# Atomic Structures of all the Twenty Essential Amino Acids and a Tripeptide, with Bond Lengths as Sums of Atomic Covalent Radii

Raji Heyrovska

Institute of Biophysics, Academy of Sciences of the Czech Republic.

Email: rheyrovs@hotmail.com

**Abstract**

Recently, the bond lengths of the molecular components of nucleic acids and of caffeine and related molecules were shown to be sums of the appropriate covalent radii of the adjacent atoms. Thus, each atom was shown to have its specific contribution to the bond length. This enabled establishing their atomic structures for the first time. In this work, the known bond lengths for amino acids and the peptide bond are similarly shown to be sums of the atomic covalent radii. Based on this result, the atomic structures of all the twenty essential amino acids and a tripeptide have been presented.

## 1. Introduction

The author has shown [1] in recent years that the lengths of the completely ionic, partially ionic and completely covalent bonds and of the hydrogen bonds in many inorganic and biochemical groups [2] can be considered as sums of the covalent and or ionic radii of the atoms or ions constituting the bonds. Most recently, the bond lengths in the molecular components of nucleic acids [3] and in caffeine and related molecules [4] were shown to be sums of the appropriate covalent radii of the adjacent atoms, and their atomic structures were presented for the first time. In this work, the main chemical bonds in amino acids [5] (see Fig. 1) and in the peptide bond have similarly been shown to be sums of the covalent radii of the adjacent atoms. The bond length of the NC peptide bond, which is known to be shorter than the N-C single bond due to some double bond character, has been interpreted here as the sum the covalent double bond radius of C and single bond radius of N. Thus, note that whereas in the literature, bond lengths are usually interpreted [6] as characteristic of the bonds between the two atoms, this author finds that, in general, bond lengths are sums of two distances characteristic for each of the atoms or ions constituting the bonds. The above results have thus enabled to establish the atomic structures of all the essential amino acids with bond lengths as sums of the appropriate radii of the adjacent atoms.

## 2. Bond lengths in amino acids as sums of the atomic covalent radii

The atomic covalent radii [6] or the bonding atomic radii [7], d(A) is a distance defined as, $R_{cov} = d(A) = d(AA)/2$, where d(AA) is the known inter-atomic distance. The atomic single bond (subscript: s.b.) covalent radius of an atom A is [6] as half of the single bond distance, d(A-A) and the double bond (subscript: d.b.) covalent radius is half of the double bond distance d(A=A). The latter is smaller than the single bond distance [6]. Fig. 2A gives the values of $R_{cov}$ for C, N, O and H atoms (data in [6], [8]), as shown under the circles of the given radii. The radius of $C_{res}$, as in graphite [6], also forms the resonance bond in benzene [1,6] and in the atomic structures of the molecular components of nuclei acids [3] and caffeine related molecules [4]. For a covalent bond between any two atoms, Pauling [6] considered the bond length as the sum of the covalent radii. However, he did not extend this to the biological molecules under consideration here.

The average values of the main bond lengths for amino acids from [6, 9-14] are presented in Table 1. The corresponding sums of the covalent radii R(sum) of the adjacent atoms in the bonds in the above molecules (see e.g., glycine, alanine and serine in Fig. 2B) are also given in Table 1 (see column 3).

Fig. 3 shows a plot of the various bond lengths [6, 9-14] in Table 1 versus the sum of the atomic covalent radii, R(sum). A least square line drawn through all the points gives a slope of 1.18 and an intercept of -0.25 as shown in Fig. 3. The calculated values of R(sum)cal, using the values of the slope and intercept, are given in the last but one column in Table 1. The last column gives the difference between R(sum) and R(sum)cal. This column shows that but for the deviation by 0.06 of the NH and OH bond lengths, all the rest are effectively in good agreement (< +/- 0.04). The NH and OH bond lengths [6, 9-14] are shorter perhaps because H is in the form of a 'transient' proton, $H^+$ due to H-bonding [2]. The radius of $H^+$ is 0.28 Å [1,2], whereas the covalent radius of H atom is 0.37 Å (see Fig. 2A).

Thus, it is concluded here that the main bond lengths in the amino acids are sums of the atomic covalent radii of the adjacent atoms. The atomic structures of glycine, alanine and serine (taken as examples) are drawn to scale in Fig. 2B. Note that this basic structure holds for any amino acid since the various amino acids differ only by the side group, R (see Fig. 1). Based on this result, the conventional structures such as those in [15] for molecules of all the essential amino acids (see Fig. 4) have now been converted into the atomic structures as shown in Fig. 5 (see pages 9-12 below), with bond lengths as the sum of the atomic radii given in Fig. 2A. All the structures refer to the undissociated neutral forms.

## 3. The length of the NC peptide bond as sum of the radii of C and N

The data on the length of the NC peptide bond are also given in the last but one row of Table 1. It has been found to be shorter than the $N-C_\alpha$ bond, which is usually attributed to some double bond character [6]. The linear graph with an effective slope of unity in Fig. 3 includes the bond length of the peptide bond. Thus it is shown here that, whereas the length of the $N-C_\alpha$ bond is the sum of the radii of $C_{s.b.}$ and $N_{s.b.}$ (= 0.77 + 0.70 = 1.47 Å), the length of the NC peptide bond is shorter since it is sum of the atomic covalent radii of $C_{d.b.}$ and $N_{s.b.}$ (= 0.67 + 0.70 Å = 1.37 Å). Fig. 2C shows the atomic structure of the tripeptide, Glycyl-alanyl-serine with the two NC peptide bonds.

**Acknowledgements:** The author thanks Prof. Emil Palecek of the Institute of Biophysics (IBP) for his appreciation of the research and the IBP for financial support. Grateful thanks are due to Dr. Stephen A. Salisbury of The Cambridge Crystallographic Data Centre (see ref 11), for kindly providing the carefully evaluated bond length data and for pointing that the data in ref [12] are still considered as standard. I thank Dr. Jindrich Hasek of IMC, Academy of Sciences of the Czech Republic and Prof. Robert Huber of Max-Planck-Institut fur Biochemie, Martinsried, Germany for the copies of the paper mentioned in [12].

**Table 1:** Bond lengths from [6, 9-14] and the corresponding sums of atomic covalent radii R(sum) in amino acids and in the N-C peptide bond. R(sum)cal values are as per the lst. sq. line in Fig. 3.

| 1 | 2 | 3 | 4 | 5 | 6 | 7 | 8 | 9 | 10 | 11 |
|---|---|---|---|---|---|---|---|---|---|---|
| Bonds | Bonds | R(sum) | [6] | [9] | [10] | [11,12] | [13] | [14] | R(sum)cal | Diffce. cols: 3-10 |
| N-H | **N$_{s.b.}$ - H** | **1.07** | 1.00 | 1.02 | 1.00 | | 1.02 | 1.01 | **1.01** | 0.06 |
| N-C$_\alpha$ | **N$_{s.b.}$ - C$_{s.b.}$** | **1.47** | 1.47 | 1.45 | 1.47 | 1.46 | 1.45 | 1.37 | **1.48** | -0.01 |
| C$_\alpha$ -H | **C$_{s.b.}$ - H** | **1.14** | 1.10 | 1.10 | 1.08 | | 1.09 | 1.08 | **1.10** | 0.04 |
| C$_\alpha$ -C | **C$_{s.b.}$ - C$_{d.b.}$** | **1.44** | 1.53 | 1.52 | 1.53 | 1.53 | 1.49 | 1.43 | **1.45** | -0.01 |
| C-O | **C$_{d.b.}$ - O$_{s.b.}$** | **1.34** | 1.36 | 1.33 | 1.36 | | | 1.38 | **1.33** | 0.01 |
| C=O | **C$_{d.b.}$ - O$_{d.b.}$** | **1.27** | 1.24 | 1.22 | 1.21 | 1.24 | 1.22 | 1.23 | **1.25** | 0.02 |
| O-H | **O$_{s.b.}$ - H** | **1.04** | 0.96 | 1.02 | 0.97 | | | 0.97 | **0.98** | 0.06 |
| **(N-C)$_{pept}$** | **N$_{s.b.}$ - C$_{d.b.}$** | **1.37** | 1.34 | | | 1.33 | 1.37 | | **1.37** | 0.00 |
| C$\alpha$ -C$_R$ | **C$_{s.b.}$ - C$_{s.b.}$** | **1.54** | 1.54 | | | 1.54 | | | **1.57** | -0.03 |
| (+/-) max | | 0.02 | 0.03 | 0.04 | 0.04 | 0.03 | 0.03 | 0.02 | | |

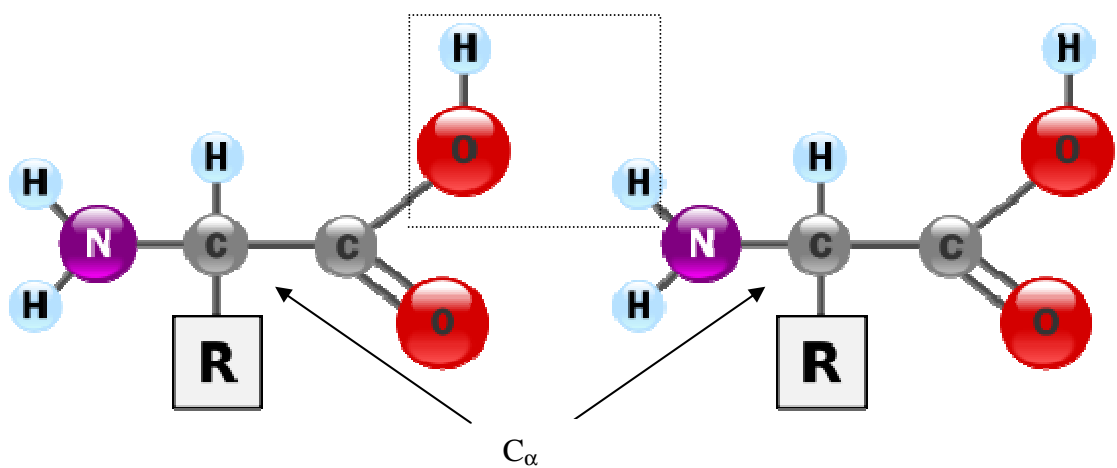

**Fig. 1**. The conventional structure and bonds in amino acids [5]. The various amino acids are distinguished by the side chain R attached to the carbon, denoted as C$_\alpha$. For example, in glycine, alanine and serine R = H, CH$_3$ and CH$_2$OH respectively. When two molecules of amino acids (shown above) condense together to form a bipeptide, a molecule of H$_2$O is formed by the combination of OH and H (as shown in the dotted rectangle) and eliminated, and an NC peptide bond is formed.

**2A.**

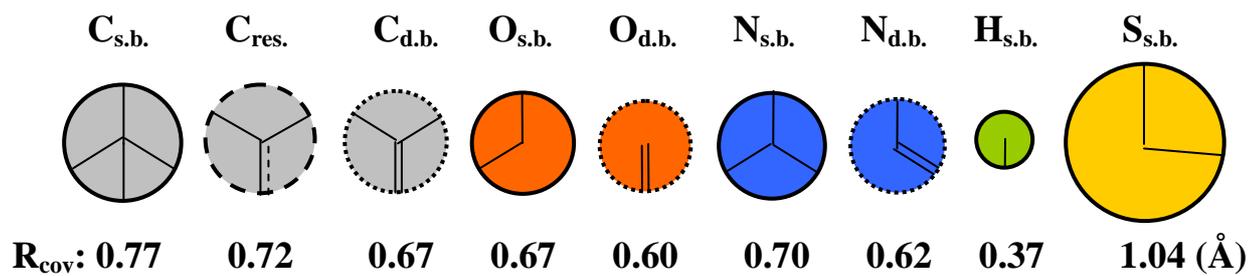

| $C_{s.b.}$ | $C_{res.}$ | $C_{d.b.}$ | $O_{s.b.}$ | $O_{d.b.}$ | $N_{s.b.}$ | $N_{d.b.}$ | $H_{s.b.}$ | $S_{s.b.}$ |

**$R_{cov}$:** 0.77   0.72   0.67   0.67   0.60   0.70   0.62   0.37   1.04 (Å)

**2B.**

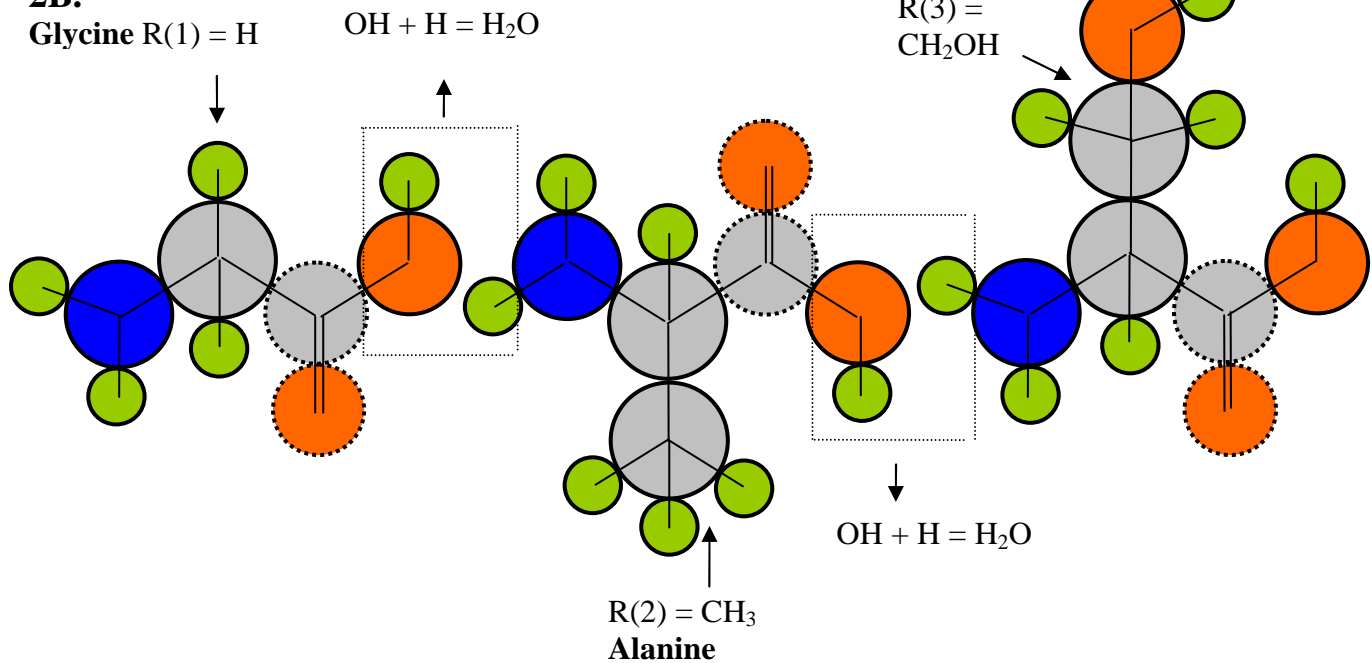

**Glycine** R(1) = H

OH + H = H$_2$O

**Serine**
R(3) =
CH$_2$OH

R(2) = CH$_3$
**Alanine**

OH + H = H$_2$O

**2C**

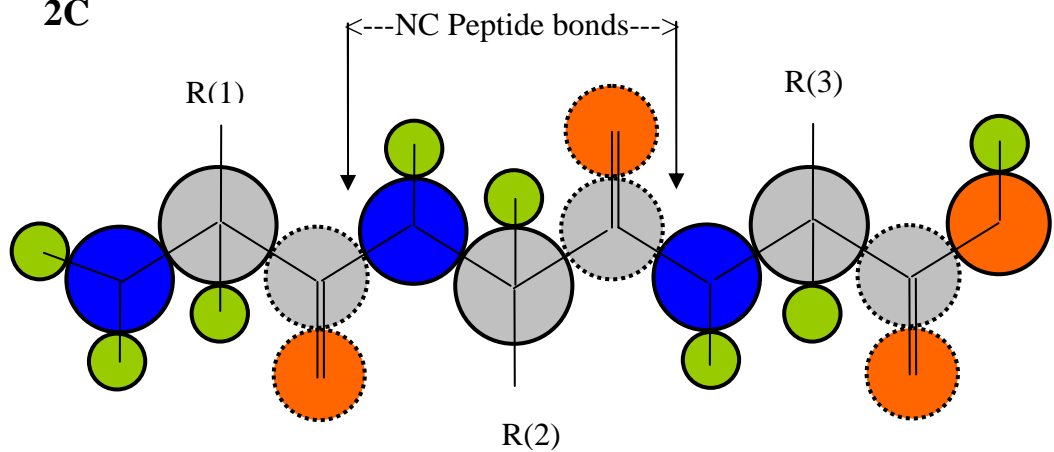

<---NC Peptide bonds--->

R(1)

R(3)

R(2)

**Fig. 2. A.** The covalent radii [6,8] of the atoms that constitute the amino acids, **B**: The three amino acids: Glycine, Alanine and Serine and **C**: their tripeptide: Glycyl-Alanyl-Serine.

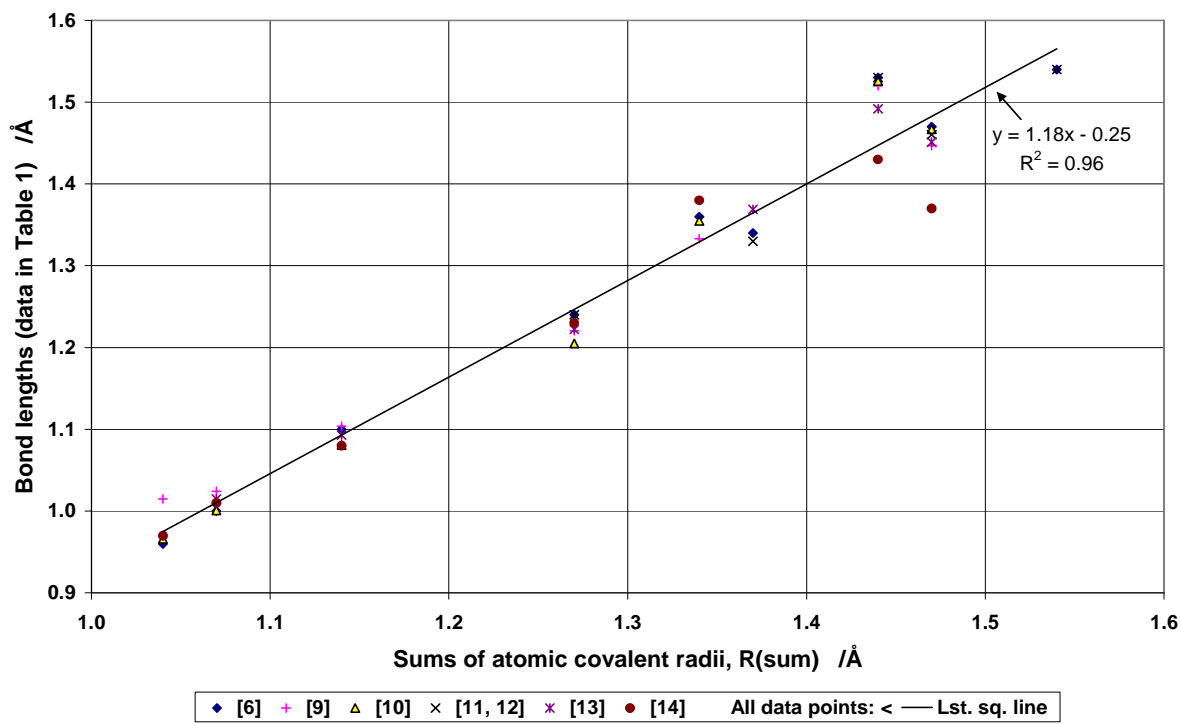

**Fig. 3.** Linear dependence of bond lengths [6, 9–14] in amino acids (see Table 1) and peptides on the corresponding sums of atomic covalent radii, R(sum) (see column 3, Table 1).

**Fig. 4.** Conventional molecular structures [15] of the "20" essential amino acids.

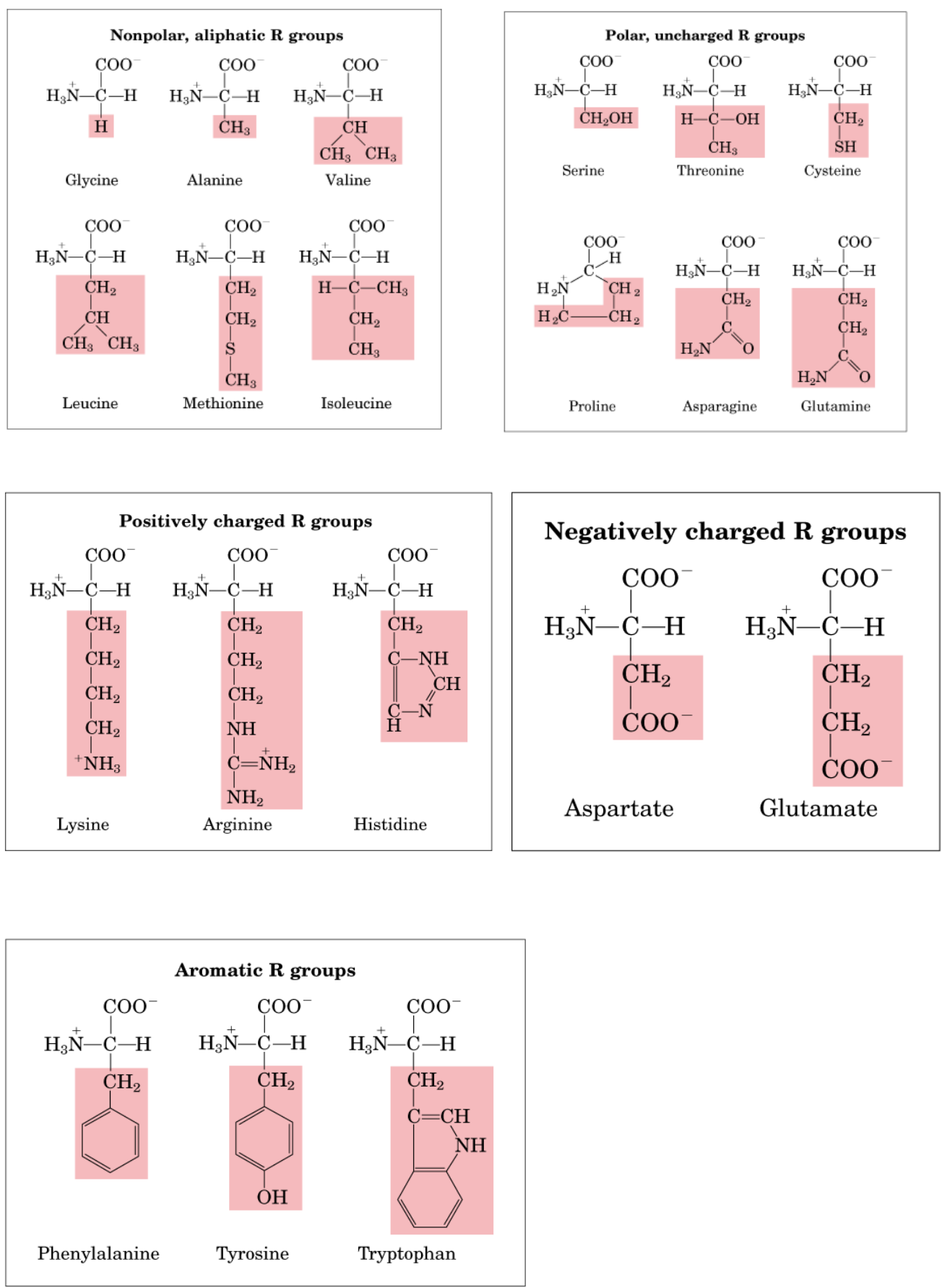

**Fig. 5. Atomic structures of "20" amino acids**

**Alanine**, R = CH₃

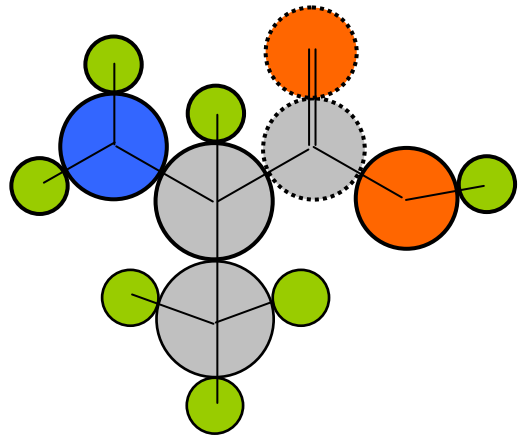

**Glycine**, R = H

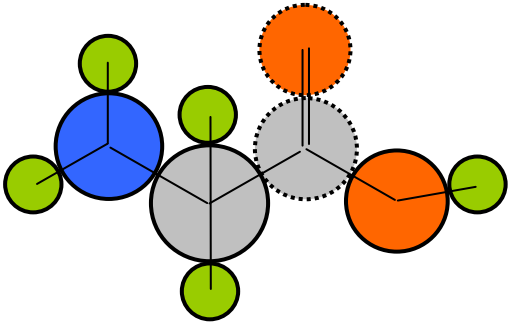

**Leucine**, R = CH₂CH(CH₃)₂

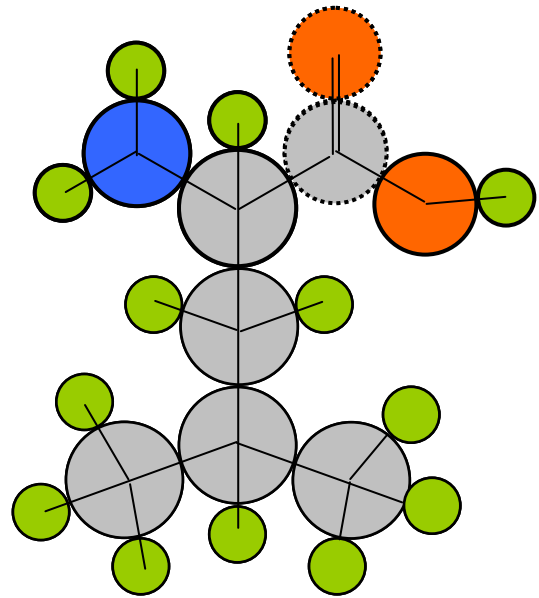

**Valine**, R = CH(CH₃)₂

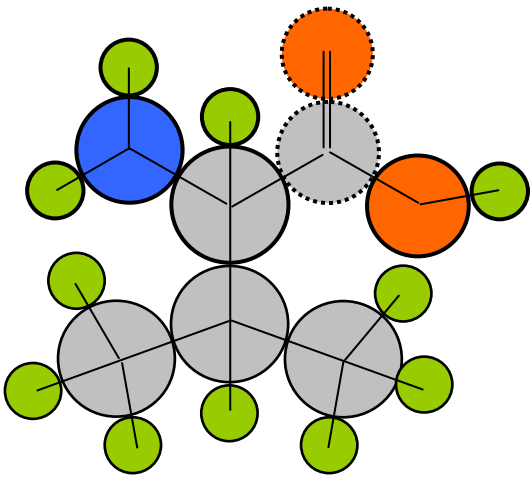

Isoleucine, R = CH(CH₃)CH₂(CH₃)

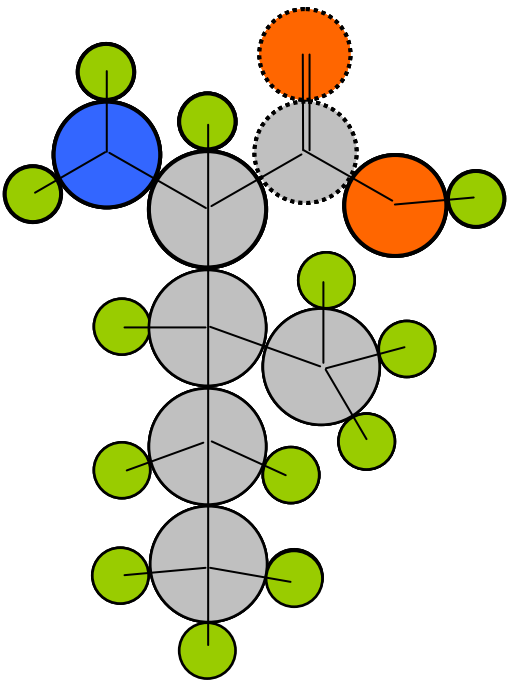

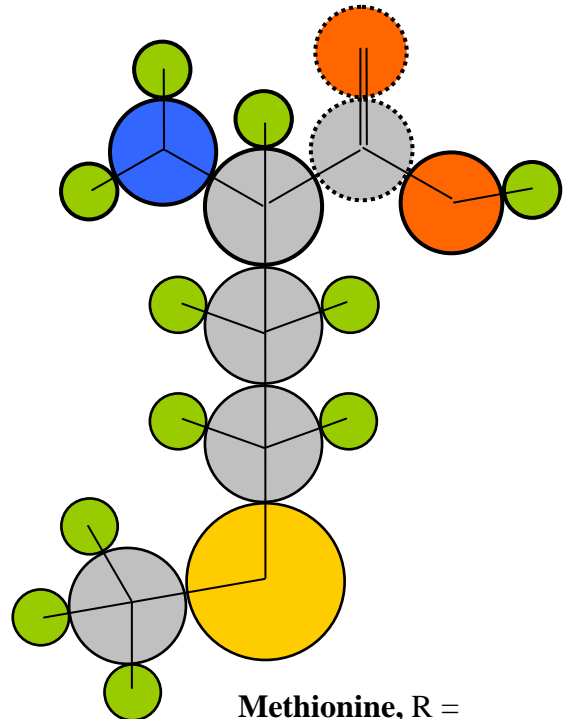

**Methionine,** R = CH₂CH₂SCH₃

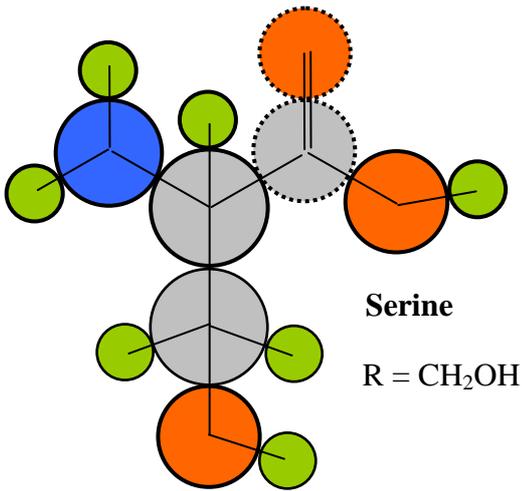

**Serine**

R = CH$_2$OH

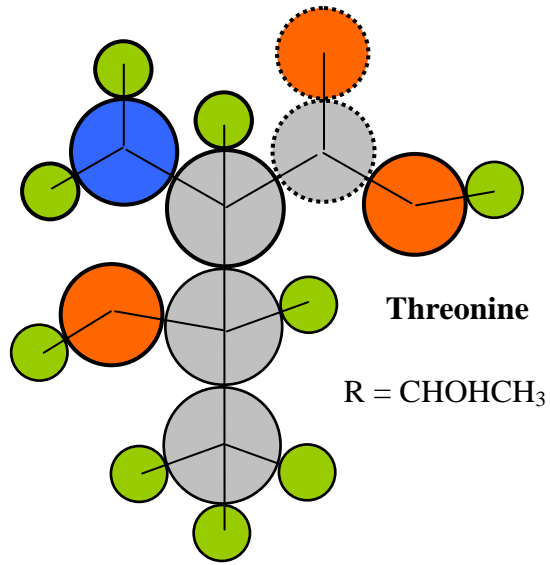

**Threonine**

R = CHOHCH$_3$

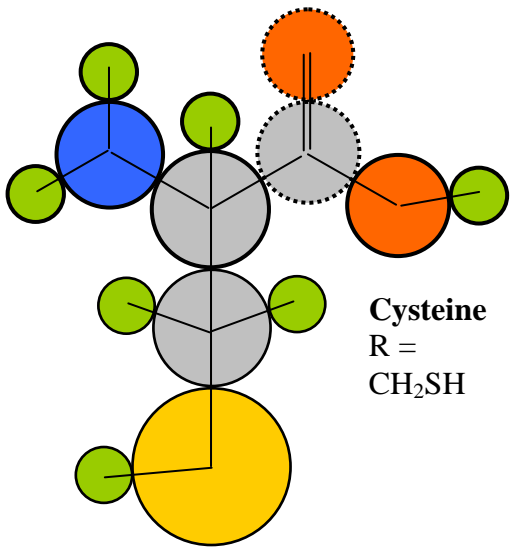

**Cysteine**
R =
CH$_2$SH

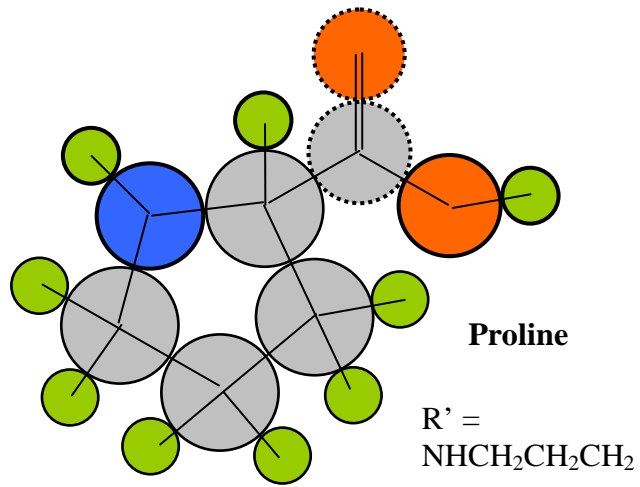

**Proline**

R' =
NHCH$_2$CH$_2$CH$_2$

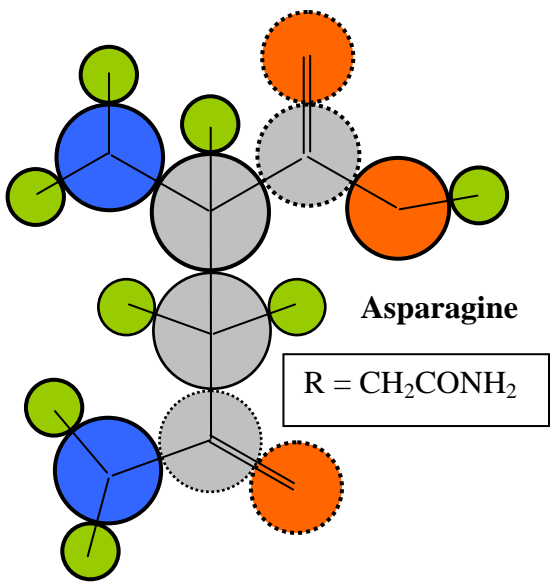

**Asparagine**

R = CH$_2$CONH$_2$

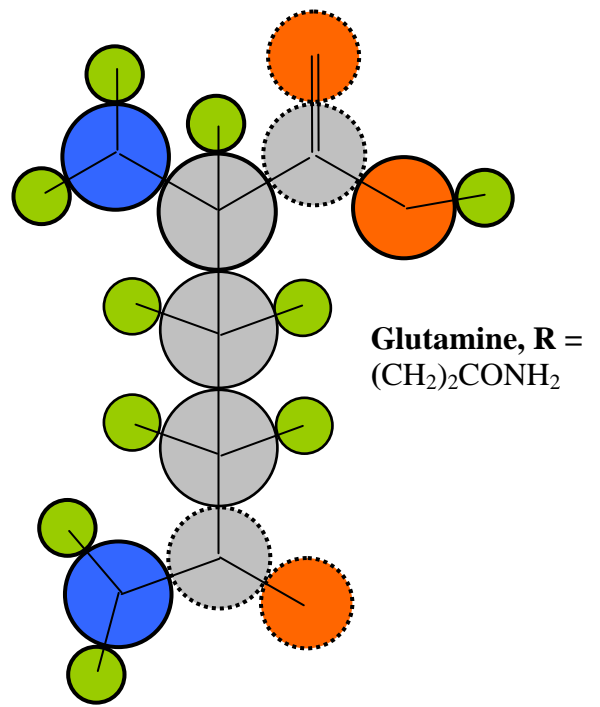

**Glutamine, R =**
(CH$_2$)$_2$CONH$_2$

**Lysine**, R = (CH$_2$)$_4$ NH$_2$

**Arginine**, R = (CH$_2$)$_3$NHC(NH)NH$_2$

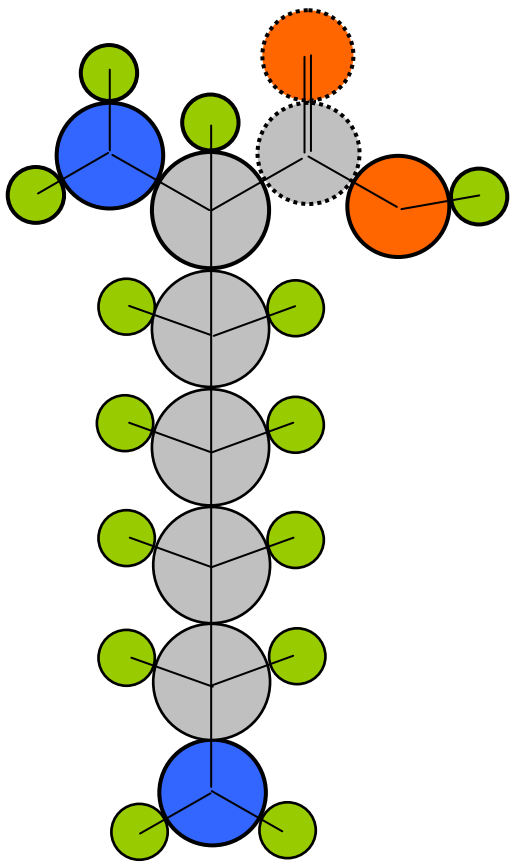

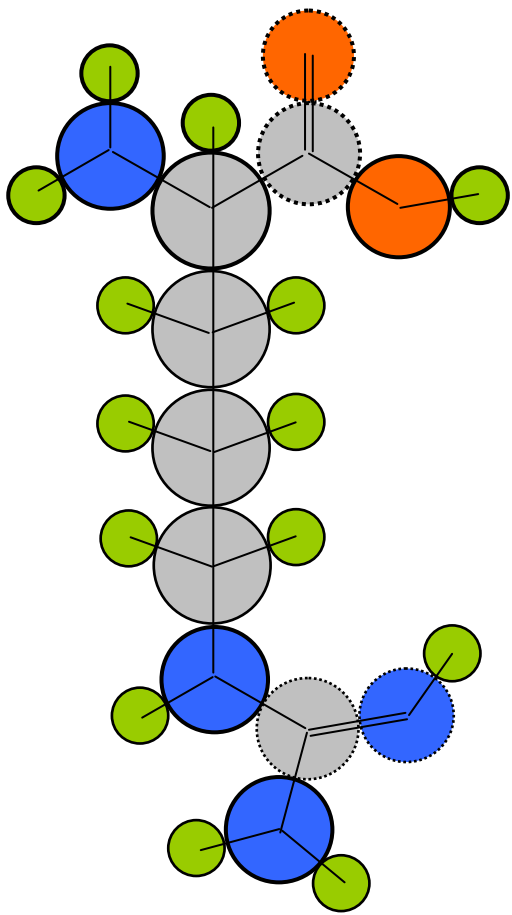

**Histidine**, R = CH$_2$(CNHCHNCH)

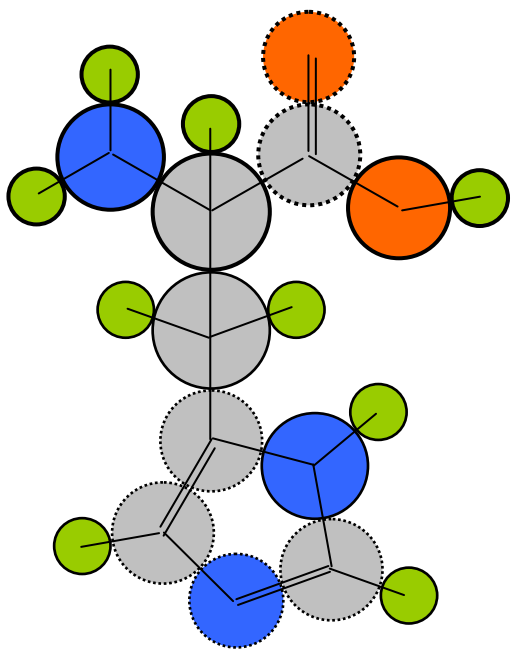

**Aspartic acid**, R = CH₂COOH     **Glutamic acid** = (CH₂)₂COOH

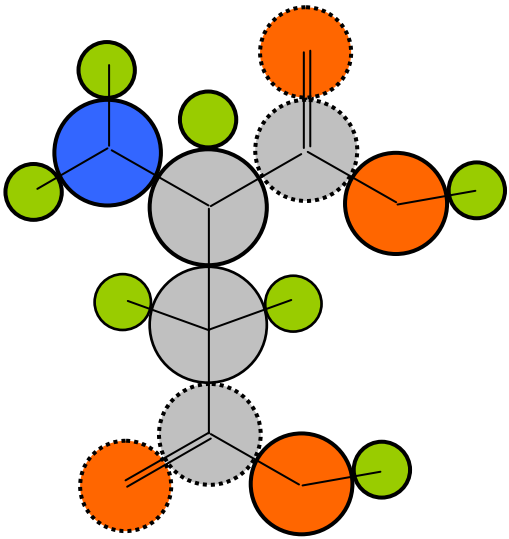

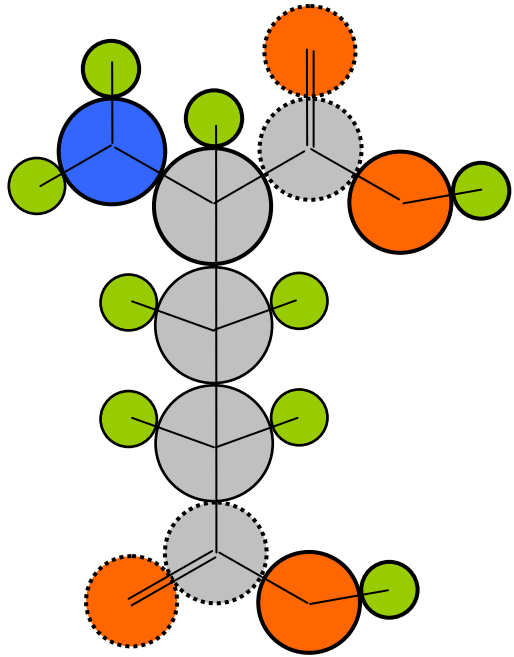

**Phenyl alanine,** R = CH₂ph

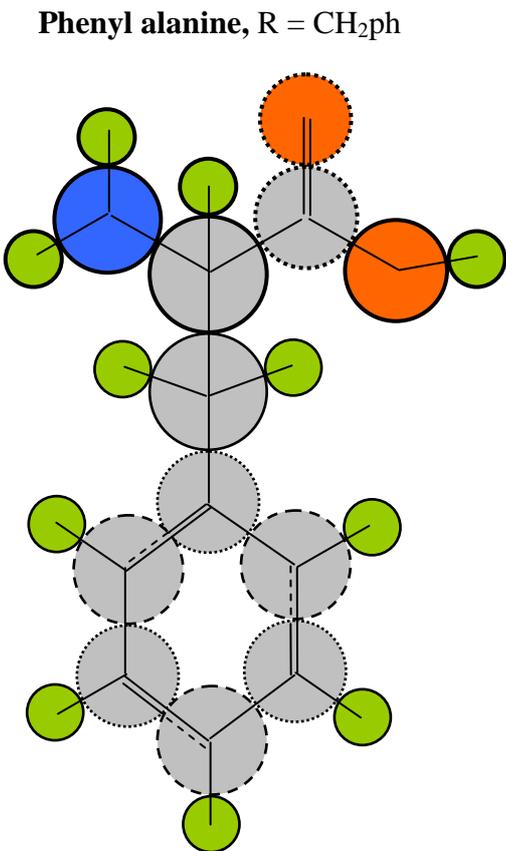

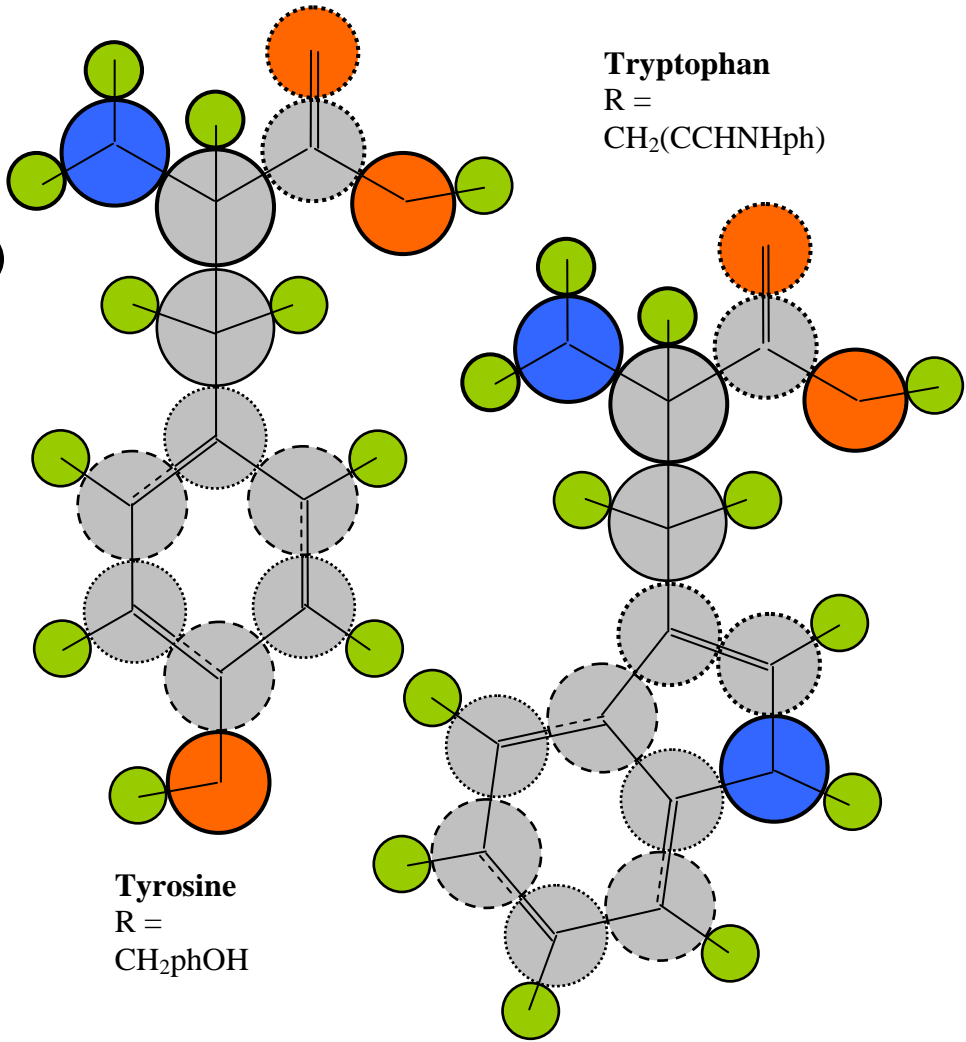

**Tryptophan**
R =
CH₂(CCHNHph)

**Tyrosine**
R =
CH₂phOH